%% file: template.tex
\documentclass[cameraready]{Interspeech}

\title{SSL-GMMVC: Interpretable Voice Conversion via Locally Linear GMM Transforms in Self-Supervised Representation Space}

\author[affiliation={1}, orcid=0009-0004-3381-8113]{Tomoya}{Tanabu}
\author[affiliation={1}, orcid=0009-0004-7524-8112]{Hiroshi}{Nishijima}
\author[affiliation={1}, orcid=0000-0002-6265-9674]{Daisuke}{Saito}
\author[affiliation={1}, orcid=0000-0002-8778-9555]{Nobuaki}{Minematsu}

\address{
  $^1$ The University of Tokyo, Japan
}

\email{\{tanabu,hiroshi,dsk\_saito,mine\}@gavo.t.u-tokyo.ac.jp}


\usepackage{comment}
\usepackage{booktabs}
\usepackage{multirow}
\usepackage{subcaption}
\usepackage{tabularx}
\usepackage{cite}
\usepackage{xurl} 
\renewcommand{\UrlFont}{\fontfamily{zi4}\selectfont}
\newcommand{\link}[2]{\href{#1}{{\UrlFont #2}}}

\begin{document}

\maketitle

\input{tex/abstract}

\input{tex/intro}

\input{tex/proposal}

\input{tex/experiment}

\input{tex/results}

\input{tex/further_analysis}

\input{tex/conclusion}

\newpage

\input{tex/ack}

\input{tex/ai_use}

\bibliographystyle{IEEEtran}
\bibliography{mybib}

\end{document}

%% file: tex/abstract.tex



\begin{abstract}
We introduce SSL-GMMVC, an interpretable voice conversion method in self-supervised speech space.
The method models paired source-target features with a Gaussian mixture model and performs conversion as a posterior-weighted sum of affine transforms.
This yields locally linear transformations that adapt to heterogeneous feature-space structure while remaining analytically tractable.
Through objective and subjective evaluations, we show that SSL-GMMVC improves speaker similarity with comparable intelligibility and naturalness,
and that even a constrained covariance variant surpasses a deep learning baseline as the number of mixture components increases.
Further analyses link component selection to phonetic structure and reveal interpretable scaling and rotation in the learned transforms.
These findings highlight SSL-GMMVC as an effective, analyzable framework for voice conversion.
\end{abstract}

%% file: tex/intro.tex
\section{Introduction}

Voice conversion (VC) modifies a speaker's voice to sound like another person
while preserving linguistic content~\cite{VC}.
Its applications include anonymization~\cite{privacy, anonymization},
computer-assisted language learning~\cite{foreign-accent, L2VC}, and speaking aid~\cite{speaking-aid, dysarthria}.

VC progress has been driven by advances in both transformation models and speech representations~\cite{OverviewVC,VCStats2Deep}.
Self-supervised learning (SSL) has yielded general-purpose acoustic embeddings,
including wav2vec 2.0~\cite{wav2vec2}, HuBERT~\cite{HuBERT}, and WavLM~\cite{WavLM},
that transfer well across various downstream tasks~\cite{AudioSSL, SSLReview, SUPERB}.
Several VC systems, such as S3PRL-VC~\cite{S3PRL-VC}, FreeVC~\cite{FreeVC}, and AdaptVC~\cite{AdaptVC},
now leverage these representations as input, yet still rely on complex neural architectures for the transformation itself.

At the same time, recent work suggests that competitive VC can arise from surprisingly simple operations on SSL features.
kNN-VC~\cite{kNN-VC} and LinearVC~\cite{LinearVC} show that nearest-neighbor replacement
or a single linear mapping in SSL space can yield intelligible speech with target-speaker resemblance.
However, a global linear transform cannot adapt to local structure
that may vary across phonetic clusters~\cite{discrete-tokens}, limiting its expressiveness.

To address this, we propose SSL-GMMVC\footnote{Code public at: \link{https://github.com/tomoya-san/ssl-gmmvc}{github.com/tomoya-san/ssl-gmmvc}}, a Gaussian Mixture Model (GMM)-based VC operating on SSL representations.
SSL-GMMVC replaces the single global mapping of LinearVC with a set of locally linear transformations,
preserving interpretability while being more expressive.
By focusing on simple, locally interpretable transformations in SSL space,
we aim to improve conversion performance and to better understand SSL embedding spaces.

%% file: tex/proposal.tex
\section{SSL-GMMVC}

\subsection{Overview}
\label{subsec:ssl-gmmvc_overview}

\begin{figure}[tbp]
  \centering
  \includegraphics[width=\columnwidth]{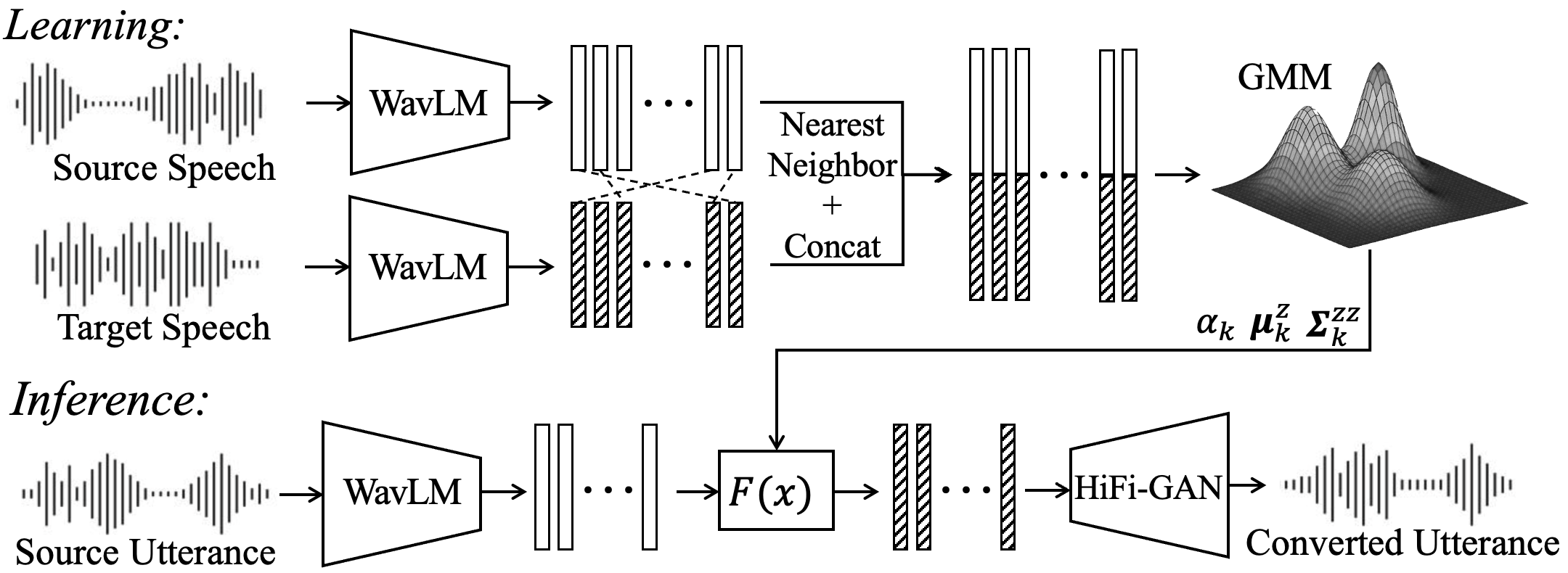}
  \caption{Overview of SSL-GMMVC}
  \label{fig:proposal}
\end{figure}

We propose SSL-GMMVC, a GMM-based voice conversion system operating on SSL features.
Figure~\ref{fig:proposal} shows an overview of the pipeline.
We align SSL features of source and target utterances using nearest-neighbor matching to obtain paired source--target vectors following~\cite{LinearVC}, and model their joint distribution with a GMM.
At conversion time, the learned parameters define an explicit function that maps source features to the target space, and a vocoder synthesizes waveforms from the converted features.

\subsection{GMM-based voice conversion}
\label{subsec:gmm_vc_ssl}

Let $\bm{x}, \bm{y}\in\mathbb{R}^{D}$ denote frame-level SSL features of the source and target speakers.
After alignment, we apply GMM-based voice conversion~\cite{GMM-original}
by fitting a $K$-component GMM to the joint vectors $\bm{z}=[\bm{x}^\top,\bm{y}^\top]^\top\in\mathbb{R}^{2D}$:
\begin{align}
  p(\bm{z}) = \sum_{k=1}^{K} \alpha_k \, \mathcal{N}(\bm{z}; \bm{\mu}_k^{z}, \bm{\Sigma}_k^{zz})
  \label{eq:gmmvc_joint_ssl}
\end{align}
where $\alpha_k$ is a mixture weight, and $\bm{\mu}_k^{z}$ and $\bm{\Sigma}_k^{zz}$ are the mean and covariance of component $k$.
We partition the mean and covariance into source and target blocks:
\begin{align}
  \bm{\mu}_k^{z} =
  \begin{bmatrix}
    \bm{\mu}_k^{x} \\
    \bm{\mu}_k^{y}
  \end{bmatrix}
  \qquad
  \bm{\Sigma}_k^{zz} =
  \begin{bmatrix}
    \bm{\Sigma}_k^{xx} & \bm{\Sigma}_k^{xy} \\
    \bm{\Sigma}_k^{yx} & \bm{\Sigma}_k^{yy}
  \end{bmatrix}
  \label{eq:mean-cov}
\end{align}

\noindent \textbf{Learning.}
We estimate $\{\alpha_k,\bm{\mu}_k^{z},\bm{\Sigma}_k^{zz}\}_{k=1}^{K}$ using the EM algorithm.
The E-step computes the responsibility $p(k|\bm{z}_n)$ for each data point $\bm{z}_n$:
\begin{align}
  p(k|\bm{z}_n) =
  \frac{\alpha_k \mathcal{N} (\bm{z}_n; \bm{\mu}_k^{z}, \bm{\Sigma}_k^{zz})}
  {\sum_{m=1}^{K}\alpha_m \mathcal{N}(\bm{z}_n; \bm{\mu}_m^{z}, \bm{\Sigma}_m^{zz})}
  \label{eq:gmmvc_resp_ssl}
\end{align}
\noindent
The M-step updates mixture weights, means, and covariances, and we iterate until convergence.

\noindent \textbf{Inference.}
Given a source feature $\bm{x}$, we compute the posterior of each component from the source-side marginal:
\begin{align}
  p(k|\bm{x}) =
  \frac{\alpha_k \mathcal{N}(\bm{x}; \bm{\mu}_k^{x}, \bm{\Sigma}_k^{xx})}
  {\sum_{m=1}^{K}\alpha_m \mathcal{N}(\bm{x}; \bm{\mu}_m^{x}, \bm{\Sigma}_m^{xx})}
  \label{eq:gmmvc_post_ssl}
\end{align}
	The converted feature is a weighted sum of component-wise affine transforms:
\begin{align}
  \label{eq:gmmvc_conversion_ssl}
  \hat{\bm{y}} {=} F(\bm{x}) {=} \sum_{k=1}^{K} p(k|\bm{x})\,\bigl\{\bm{\mu}_k^y + \bm{\Sigma}_k^{yx}(\bm{\Sigma}_k^{xx})^{-1}(\bm{x} - \bm{\mu}_k^x)\bigr\}
\end{align}

\subsection{Mathematical relationship between SSL-GMMVC and LinearVC}
\label{subsec:math_equivalence}

SSL-GMMVC can be viewed as an extension of LinearVC that replaces a single global mapping with a mixture of locally linear mappings.
Setting $K=1$ in (\ref{eq:gmmvc_conversion_ssl}) gives $p(k=1|\bm{x})=1$, reducing the conversion to the affine transform:
\begin{align}
  F(\bm{x}) = \bm{\mu}^y + \bm{\Sigma}^{yx} (\bm{\Sigma}^{xx})^{-1} (\bm{x} - \bm{\mu}^x)
  \label{eq:affine}
\end{align}
Letting $\bm{W} = (\bm{\Sigma}^{xx})^{-1}\bm{\Sigma}^{xy}$ and $\bm{b} = \bm{\mu}^y - \bm{W}^\top \bm{\mu}^x$ gives $F(\bm{x}) = \bm{W}^\top \bm{x} + \bm{b}$, which is mathematically equivalent to LinearVC.
For $K>1$, each component applies a different affine transform, yielding a locally linear but globally nonlinear mapping that captures local structure.

%% file: tex/experiment.tex
\section{Voice conversion experiments}

\subsection{Data}
\label{subsec:data}

We used American English speech from CMU ARCTIC~\cite{CMU-ARCTIC}, selecting three male (bdl, rms, aew) and three female (slt, clb, lnh) speakers.
Each utterance was around 2--3 seconds long.

\subsection{Implementation of SSL-GMMVC}

Following~\cite{kNN-VC, LinearVC}, we extracted 1024-dimensional SSL features from the 6th layer of WavLM-Large~\cite{WavLM},
which mainly captures speaker information, using \SI{20}{\milli\second} frames from \SI{16}{\kilo\hertz} audio.
We aligned source and target features via bidirectional cosine-similarity nearest-neighbor matching.
GMM training followed section~\ref{subsec:gmm_vc_ssl} with two covariance variants: \textit{Full} (\textit{F}), where the covariance matrix is unconstrained, and \textit{Cross Diag} (\textit{CD}), where all four blocks in (\ref{eq:mean-cov}) are diagonal.
\textit{CD} was widely used in traditional GMM-VC to prevent overfitting and was included here to examine model complexity effects.
Waveforms were synthesized with a HiFi-GAN~\cite{HiFi-GAN} vocoder trained by~\cite{kNN-VC} to accept WavLM-Large 6th-layer features\footnote{\link{https://github.com/bshall/knn-vc}{github.com/bshall/knn-vc}}.

\subsection{VC models for evaluation}

We compared SSL-GMMVC variants and baselines across training-set sizes of $N {\in} \{10, 20, 50, 100, 200, 300\}$ utterances.
SSL-GMMVC spanned six configurations combining covariance types \textit{F}/\textit{CD} and the number of mixtures $K {\in} \{1, 2, 4\}$.
Since larger $K$ increases the total number of parameters to estimate, which becomes unstable with limited data, we capped $K$ at 4 and required $N {\geq} 50$ for $K{=}2$ and $N {\geq} 100$ for $K{=}4$ for stable estimation.
For LinearVC, we evaluated \textit{NC} (\textit{No Constraint}; unconstrained affine transform) and \textit{BO} (\textit{Bias Only}; per-dimension mean shift).
As a deep baseline, we used FreeVC~\cite{FreeVC}, a VITS-based zero-shot model. It is the most comparable model to our approach, as it uses WavLM-Large features and HiFi-GAN synthesis.
In preliminary experiments, increasing the number of reference utterances beyond 10 did not improve speaker similarity, so we used 10 throughout.

\subsection{Objective evaluation}

\noindent \textbf{Setup.}
For all 30 ordered pairs from 6 speakers, we trained each model and evaluated converted speech in terms of speaker similarity, intelligibility and naturalness.
Intelligibility and naturalness were evaluated on 40 utterances not used for training.

\noindent \textbf{Speaker similarity.}
We computed equal error rate (EER) using a speaker verifier that scores ECAPA-TDNN~\cite{ECAPA-TDNN} embeddings by cosine similarity\footnote{\link{https://huggingface.co/speechbrain/spkrec-ecapa-voxceleb}{huggingface.co/speechbrain/spkrec-ecapa-voxceleb}}.
We evaluated discrimination between 100 converted--real pairs (converted speech vs.\ real target speech) and 100 real--real pairs, constructed with different linguistic content.
Higher EER (up to 50\%) indicates higher similarity to the target speaker.

\noindent \textbf{Intelligibility.}
We transcribed converted speech with Whisper-base\footnote{\link{https://github.com/openai/whisper}{github.com/openai/whisper}}~\cite{Whisper} and reported word error rate (WER) against the ground-truth.
Lower WER indicates higher intelligibility.

\noindent \textbf{Naturalness.}
We estimated perceived naturalness using UTMOS~\cite{UTMOS}, a mean opinion score (MOS) prediction model.

\subsection{Subjective evaluation}
\label{subsec:sub_eval}

\noindent \textbf{Setup.}
We conducted listening tests on four speaker pairs covering all gender directions (bdl$\to$rms, clb$\to$slt, bdl$\to$slt, and clb$\to$rms).
We converted five utterances per model for each pair and collected five ratings per sample via the crowdsourcing platform Lancers.
We report the mean opinion scores (MOS) of speaker similarity and naturalness.

\noindent \textbf{Speaker similarity.}
Raters scored similarity to the target speaker on a 4-point scale.
Each trial presented a reference utterance (real target speech) and an evaluation utterance with different linguistic content, followed by a rating.
Each rater evaluated 21 trials in random order.
As an attention check, we included one real--real same-speaker trial and excluded raters who rated it as 1.

\noindent \textbf{Naturalness.}
Raters scored naturalness on a 5-point scale (5 = most natural) by listening to each utterance in isolation.
Each rater evaluated 21 samples in random order using the same attention check as for speaker similarity.

%% file: tex/results.tex
\section{Results}

\begin{table*}[tbp]
    \centering
    \caption{Objective evaluation results of speaker similarity, intelligibility and naturalness. $N$ is the number of training utterances. Underline indicates SSL-GMMVC values that surpass FreeVC. Bold indicates SSL-GMMVC \textit{F} with $K{=}2$ or $K{=}4$ matching or exceeding LinearVC \textit{NC} at the same $N$.}
    \label{tab:obj_results}
    \setlength{\tabcolsep}{1.0pt}
    \renewcommand{\arraystretch}{1.05}
    \begin{tabular*}{\textwidth}{@{\extracolsep{\fill}}@{}l l c cccccc cccccc cccccc@{}}
            \toprule
            \multicolumn{3}{c}{Model} & \multicolumn{6}{c}{EER [\%] ($\uparrow$)} & \multicolumn{6}{c}{WER [\%] ($\downarrow$)} & \multicolumn{6}{c}{UTMOS ($\uparrow$)} \\
            \cmidrule(lr){1-3}\cmidrule(lr){4-9}\cmidrule(lr){10-15}\cmidrule(lr){16-21}
            \multicolumn{3}{c}{} & \multicolumn{6}{c}{$N$} & \multicolumn{6}{c}{$N$} & \multicolumn{6}{c}{$N$}\\
            & & $K$ & 10 & 20 & 50 & 100 & 200 & 300 & 10 & 20 & 50 & 100 & 200 & 300 & 10 & 20 & 50 & 100 & 200 & 300 \\
            \midrule
                \multirow{6}{*}{\shortstack[l]{SSL-\\GMMVC}} & \multirow{3}{*}{\textit{F}} & 1 & \underline{9.58} & \underline{22.73} & \underline{26.40} & \underline{26.42} & \underline{25.78} & \underline{25.77} & 12.14 & \underline{3.64} & \underline{2.91} & \underline{2.71} & \underline{2.63} & \underline{2.70} & 3.08 & 4.12 & \underline{4.28} & \underline{4.32} & \underline{4.33} & \underline{4.33} \\
                & & 2 & -- & -- & \underline{25.00} & \underline{\textbf{26.47}} & \underline{\textbf{27.27}} & \underline{\textbf{26.37}} & -- & -- & \underline{3.06} & \underline{2.98} & \underline{2.81} & \underline{2.85} & -- & -- & 4.23 & \underline{4.31} & \underline{\textbf{4.33}} & \underline{\textbf{4.33}} \\
        & & 4 & -- & -- & -- & \underline{26.02} & \underline{\textbf{27.30}} & \underline{\textbf{27.35}} & -- & -- & -- & \underline{3.03} & \underline{2.95} & \underline{2.79} & -- & -- & -- & \underline{4.26} & \underline{4.31} & \underline{\textbf{4.33}} \\
        & \multirow{3}{*}{\textit{CD}} & 1 & 2.00 & 2.07 & 1.77 & 1.67 & 1.57 & 1.73 & \underline{3.10} & \underline{3.16} & \underline{3.13} & \underline{3.08} & \underline{3.05} & \underline{3.00} & 4.01 & 4.03 & 4.06 & 4.07 & 4.08 & 4.08 \\
        & & 2 & -- & -- & \underline{3.27} & \underline{2.97} & \underline{3.07} & \underline{2.92} & -- & -- & \underline{3.10} & \underline{3.01} & \underline{3.04} & \underline{3.12} & -- & -- & 4.07 & 4.09 & 4.10 & 4.11 \\
        & & 4 & -- & -- & -- & \underline{4.40} & \underline{4.20} & \underline{4.22} & -- & -- & -- & \underline{3.18} & \underline{3.00} & \underline{3.07} & -- & -- & -- & 4.11 & 4.13 & 4.13 \\
        \midrule
        \multirow{2}{*}{\shortstack[l]{LinearVC}} & \textit{NC} & -- & 9.58 & 22.77 & 26.45 & 26.40 & 25.88 & 25.73 & 12.20 & 3.71 & 2.93 & 2.80 & 2.61 & 2.70 & 3.08 & 4.12 & 4.28 & 4.32 & 4.33 & 4.33 \\
        & \textit{BO} & -- & 0.73 & 0.62 & 0.65 & 0.73 & 0.67 & 0.62 & 3.24 & 3.25 & 3.24 & 3.25 & 3.22 & 3.28 & 4.07 & 4.07 & 4.08 & 4.08 & 4.08 & 4.08 \\
        \midrule
        \shortstack[l]{FreeVC} & -- & -- & \multicolumn{6}{c}{2.85} & \multicolumn{6}{c}{3.85} & \multicolumn{6}{c}{4.25} \\
        \bottomrule
    \end{tabular*}%
\end{table*}

\subsection{Results of the objective evaluation}
\label{subsec:objective_results}
Table~\ref{tab:obj_results} summarizes the objective evaluation results.
Because FreeVC is pretrained, its performance is independent of training-set size.
We refer to SSL-GMMVC \textit{F} and LinearVC \textit{NC} collectively as \textit{unconstrained} models, and SSL-GMMVC \textit{CD} and LinearVC \textit{BO} as \textit{constrained} models.

\noindent\textbf{Speaker similarity.}
The EER of SSL-GMMVC \textit{F} increased steadily with training data size,
surpassing LinearVC \textit{NC} at $N {\geq} 100$ for $K{=}2$ and at $N {\geq} 200$ for $K{=}4$.
This confirms that more mixture components enable the model to more effectively exploit larger training sets.
SSL-GMMVC \textit{CD} consistently outperformed LinearVC \textit{BO} across all configurations,
despite both resulting in comparatively low EERs.
This advantage stems from \textit{CD} learning both scaling and shifting, while \textit{BO} is limited to shifting alone.
For \textit{CD}, EER rose by 1.0--1.5\% per step as $K$ increased, confirming that the increase in the number of mixtures yields consistent gains even under covariance constraints.
Overall, all SSL-GMMVC models except \textit{CD} at $K{=}1$ achieved higher EER than FreeVC.

\noindent\textbf{Intelligibility.}
\textit{Unconstrained} models showed elevated WER at $N {=} 10$ but fell below FreeVC at $N {\geq} 20$.
SSL-GMMVC \textit{F} did not surpass LinearVC \textit{NC} in intelligibility, but remained closely comparable.
\textit{Constrained} models remained intelligible even with $N {=} 10$, with SSL-GMMVC \textit{CD} outperforming LinearVC \textit{BO} across all settings.
This likely reflects their smaller parameterization, which learns content-preserving mappings more reliably from limited data while limiting speaker resemblance.

\noindent\textbf{Naturalness.}
In the \textit{unconstrained} setting, SSL-GMMVC \textit{F} with $K{>}1$ reached naturalness on par with LinearVC \textit{NC} and above FreeVC at $N {\geq} 200$,
confirming that the GMM transformation preserves the naturalness of SSL features.
In the \textit{constrained} setting, SSL-GMMVC \textit{CD} with $K{>}1$ consistently surpassed LinearVC \textit{BO} at $N {\geq} 100$,
and even \textit{CD} at $K{=}1$ maintained stable naturalness across all training-set sizes.

\subsection{Results of the subjective evaluation}
\label{subsec:subjective_results}

\begin{table*}[tbp]
    \centering
    \caption{Subjective evaluation results of speaker similarity and naturalness (MOS with 95\% confidence interval in parentheses). $N$ indicates the number of training utterances. Underline indicates SSL-GMMVC values that surpass FreeVC. Bold indicates SSL-GMMVC \textit{F} with $K{=}2$ or $K{=}4$ matching or exceeding LinearVC \textit{NC} at the same $N$.}
    \label{tab:mos_results}
    \setlength{\tabcolsep}{1.0pt}
    \renewcommand{\arraystretch}{1.05}
    \begin{tabular*}{\textwidth}{@{\extracolsep{\fill}}@{}l l c cccccc cccccc@{}}
        \toprule
            \multicolumn{3}{c}{Model} & \multicolumn{6}{c}{Speaker similarity ($\uparrow$)} & \multicolumn{6}{c}{Naturalness ($\uparrow$)} \\
            \cmidrule(lr){1-3}\cmidrule(lr){4-9}\cmidrule(lr){10-15}
            \multicolumn{3}{c}{} & \multicolumn{6}{c}{$N$} & \multicolumn{6}{c}{$N$}\\
            & & $K$ & 10 & 20 & 50 & 100 & 200 & 300 & 10 & 20 & 50 & 100 & 200 & 300 \\
            \midrule
        \multirow{6}{*}{\shortstack[l]{SSL-\\GMMVC}} & \multirow{3}{*}{\textit{F}} & 1 & 1.91(18) & \underline{2.42(19)} & \underline{2.64(20)} & \underline{2.76(19)} & \underline{2.97(19)} & \underline{2.84(21)} & 2.51(22) & 3.75(23) & 4.00(21) & 3.94(21) & 3.99(20) & 4.07(19) \\
        & & 2 & -- & -- & \underline{\textbf{2.67}(20)} & \underline{2.80(17)} & \underline{\textbf{2.88}(18)} & \underline{\textbf{2.90}(19)} & -- & -- & 3.92(19) & \textbf{4.00}(18) & 3.97(18) & 4.09(19) \\
        & & 4 & -- & -- & -- & \underline{2.74(19)} & \underline{2.65(19)} & \underline{\textbf{2.85}(18)} & -- & -- & -- & \textbf{4.01}(20) & 3.96(18) & 4.10(18) \\
        & \multirow{3}{*}{\textit{CD}} & 1 & 2.02(20) & 1.84(19) & 2.00(19) & 1.87(19) & 1.93(18) & \underline{2.05(22)} & 3.70(21) & 3.66(20) & 3.61(20) & 3.72(22) & 3.93(20) & 3.84(20) \\
        & & 2 & -- & -- & \underline{2.17(21)} & \underline{2.25(18)} & \underline{2.17(18)} & 2.03(19) & -- & -- & 3.74(20) & 3.60(20) & 3.82(20) & 3.85(19) \\
        & & 4 & -- & -- & -- & \underline{2.29(21)} & \underline{2.35(21)} & \underline{2.19(21)} & -- & -- & -- & 3.79(20) & 3.46(20) & 3.94(21) \\
            \midrule
        \multirow{2}{*}{\shortstack[l]{LinearVC}} & \textit{NC} & -- & 1.59(16) & 2.45(20) & 2.63(20) & 2.87(20) & 2.75(20) & 2.64(19) & 2.33(22) & 3.74(20) & 4.02(20) & 3.90(20) & 4.05(19) & 4.21(18) \\
        & \textit{BO} & -- & 1.71(18) & 1.64(16) & 1.84(20) & 1.73(18) & 1.58(18) & 1.58(16) & 3.72(21) & 3.70(22) & 3.68(23) & 3.81(22) & 3.54(24) & 3.77(22) \\
        \midrule
        \shortstack[l]{FreeVC} & -- & -- & \multicolumn{6}{c}{2.04(20)} & \multicolumn{6}{c}{4.11(18)} \\
        \bottomrule
    \end{tabular*}%
\end{table*}

Table~\ref{tab:mos_results} reports the subjective evaluation results.

\noindent\textbf{Speaker similarity.}
With $N {\geq} 20$, all \textit{unconstrained} models surpassed FreeVC, and MOS increased with larger training data.
SSL-GMMVC \textit{F} exceeded LinearVC \textit{NC} at $N {\geq} 200$, broadly consistent with the objective results.
In the \textit{constrained} settings, SSL-GMMVC \textit{CD} consistently surpassed LinearVC \textit{BO} and improved with larger $K$, aligning with the objective findings.
At $N {=} 10$, however, \textit{unconstrained} models fell below FreeVC
due to conversion artifacts from insufficient data.

\noindent\textbf{Naturalness.}
\textit{Unconstrained} models showed low naturalness at $N {=} 10$ due to conversion artifacts, but recovered sharply with $N {\geq} 20$ and approached FreeVC-level naturalness as training-set size increased.
\textit{Constrained} models, by contrast, maintained stable naturalness across all training-set sizes, consistent with the objective results.

%% file: tex/further_analysis.tex
\section{Further analysis}
This section investigates how mixture components relate to phonetic categories
and how the transformation matrices act on the feature space across speaker pairs.

\subsection{Component selection}

\subsubsection{Setup}
We analyzed SSL-GMMVC \textit{F} ($K{=}2$, $N {=} 200$) with 50 held-out utterances per speaker pair.
Frames were labeled as sonorants or obstruents~\cite[Chapter~5]{phonological-analysis} via Montreal Forced Aligner~\cite{mfa}, and purity~\cite{purity} was computed using posteriors (\ref{eq:gmmvc_post_ssl}) as soft assignments, where $\mathcal{C}=\{\text{son},\text{obs}\}$ and $T$ is the number of non-silent frames:
\begin{align}
  \text{Purity}=\frac{1}{T}\sum_{k}\max_{c\in\mathcal{C}}\sum_{t:\,c_t=c} p(k|\bm{x}_t)
  \label{eq:purity}
\end{align}

\subsubsection{Results}
\begin{figure}[tbp]
    \centering
    \includegraphics[width=\columnwidth]{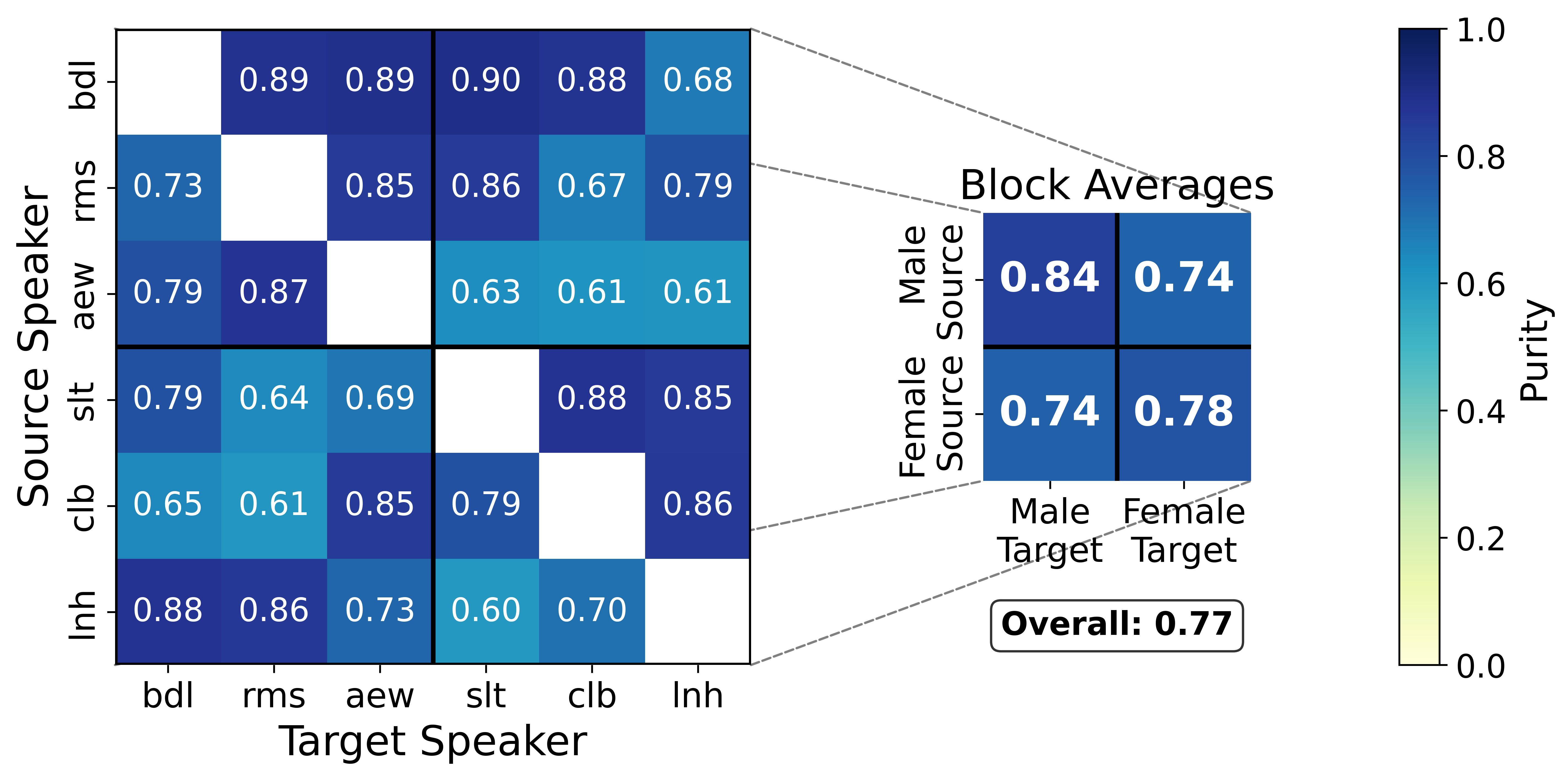}
    \caption{Purity between mixture selection and sonority (left: speaker pairs; right: gender-pair averages).}
    \label{fig:purity}
\end{figure}

Figure~\ref{fig:purity} showed a relatively high overall purity, indicating the correlation between mixture selection and sonority,
though variation existed across pairs.
More specifically, same-gender conversions tended to achieve higher purity than cross-gender ones.
\subsection{Transformation matrices}

\subsubsection{Setup}
Using the same evaluation set, we analyzed SSL-GMMVC \textit{F} ($K{=}1$, $N{=}200$).
We computed mean per-frame cosine angles between source and converted SSL features across all, sonorant, and obstruent regions.
However, identical angles can correspond to very different directions in 1024-dimensional space, so angles alone cannot reveal the global transformation structure.
We therefore additionally analyzed the eigenvalues of the conversion matrix $\bm{W}$ from subsection~\ref{subsec:math_equivalence}.
For an eigenvalue $\lambda=r e^{i\theta}$, $r$ captures scaling and $\theta$ captures the angle of the corresponding rotational plane~\cite{rotational}.
Although our primary contribution is the multi-component formulation, we restricted this analysis to $K{=}1$.
For $K{>}1$, quantitative comparison of per-component spectra is not yet established, as it requires matching rotational planes across components, for which no principled method exists.
Thus, we provided an interpretable characterization of the general transformation using a single-component model.

\subsubsection{Results}
\begin{figure}[tbp]
    \centering
    \includegraphics[width=\columnwidth]{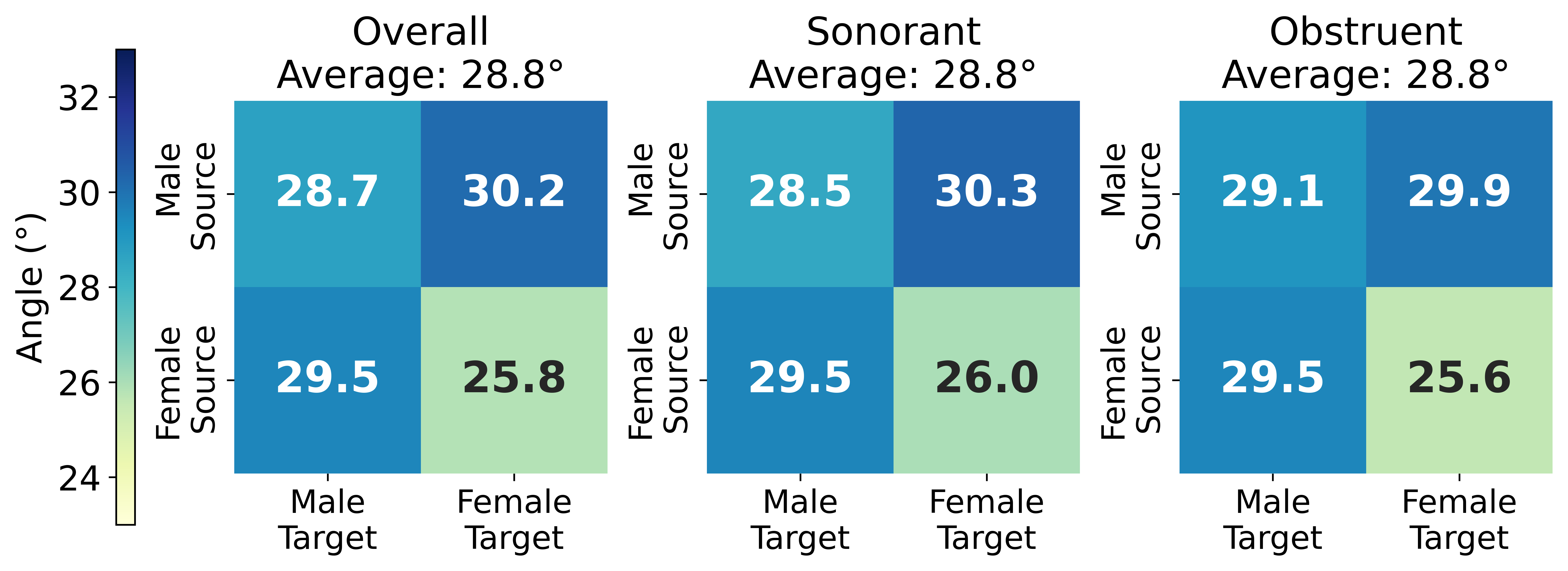}
    \caption{Average per-frame cosine angles between source and converted SSL features reported for each gender direction and for all / sonorant / obstruent regions.}
    \label{fig:angles}
\end{figure}

\begin{figure}[tbp]
    \centering
    \begin{minipage}[b]{0.49\columnwidth}
        \centering
        \includegraphics[width=\columnwidth]{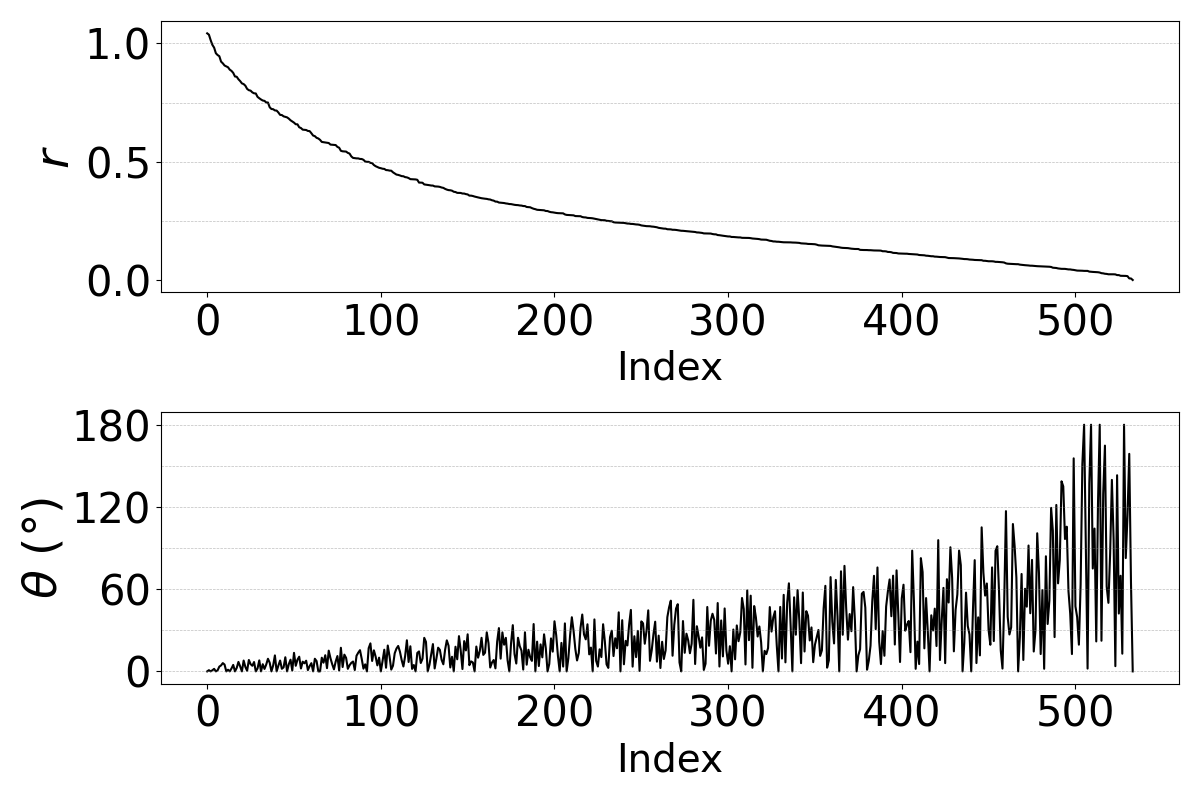}
        (a) bdl $\to$ rms
    \end{minipage}
    \hfill
    \begin{minipage}[b]{0.49\columnwidth}
        \centering
        \includegraphics[width=\columnwidth]{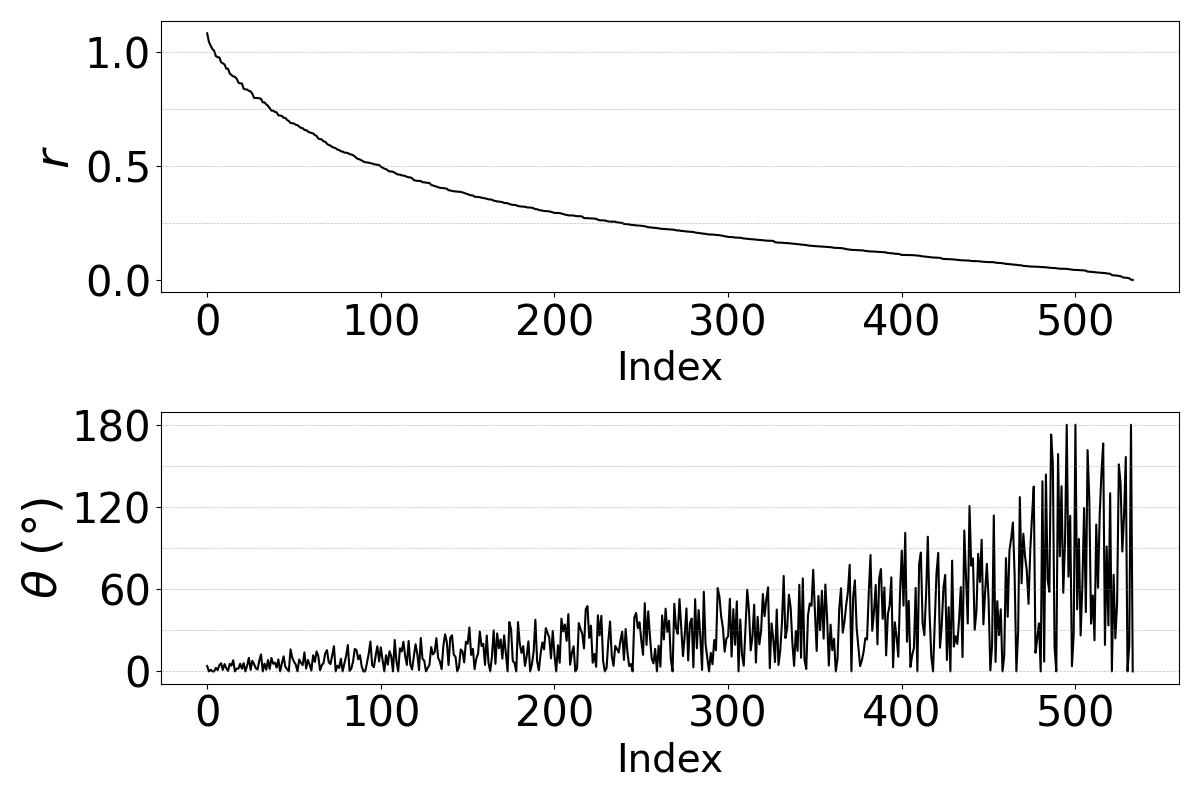}
        (b) bdl $\to$ slt
    \end{minipage}
    \\[1em]
    \begin{minipage}[b]{0.49\columnwidth}
        \centering
        \includegraphics[width=\columnwidth]{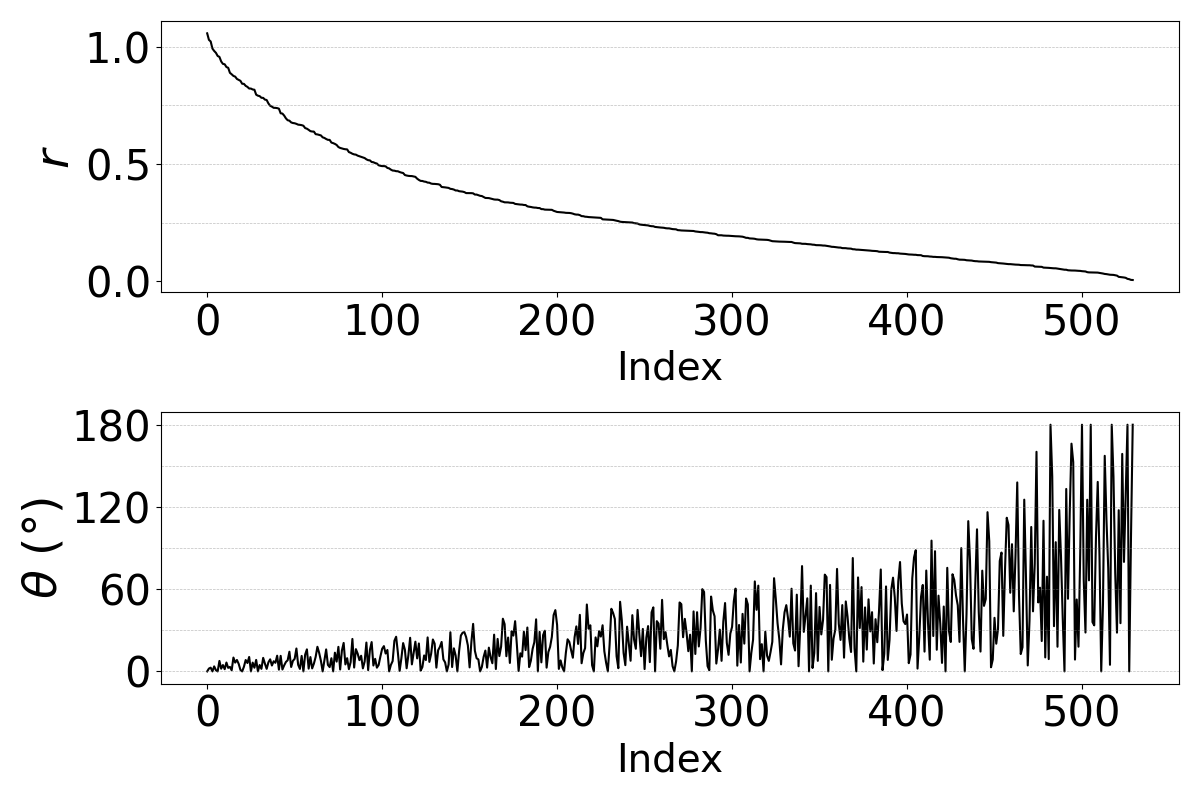}
        (c) clb $\to$ rms
    \end{minipage}
    \hfill
    \begin{minipage}[b]{0.49\columnwidth}
        \centering
        \includegraphics[width=\columnwidth]{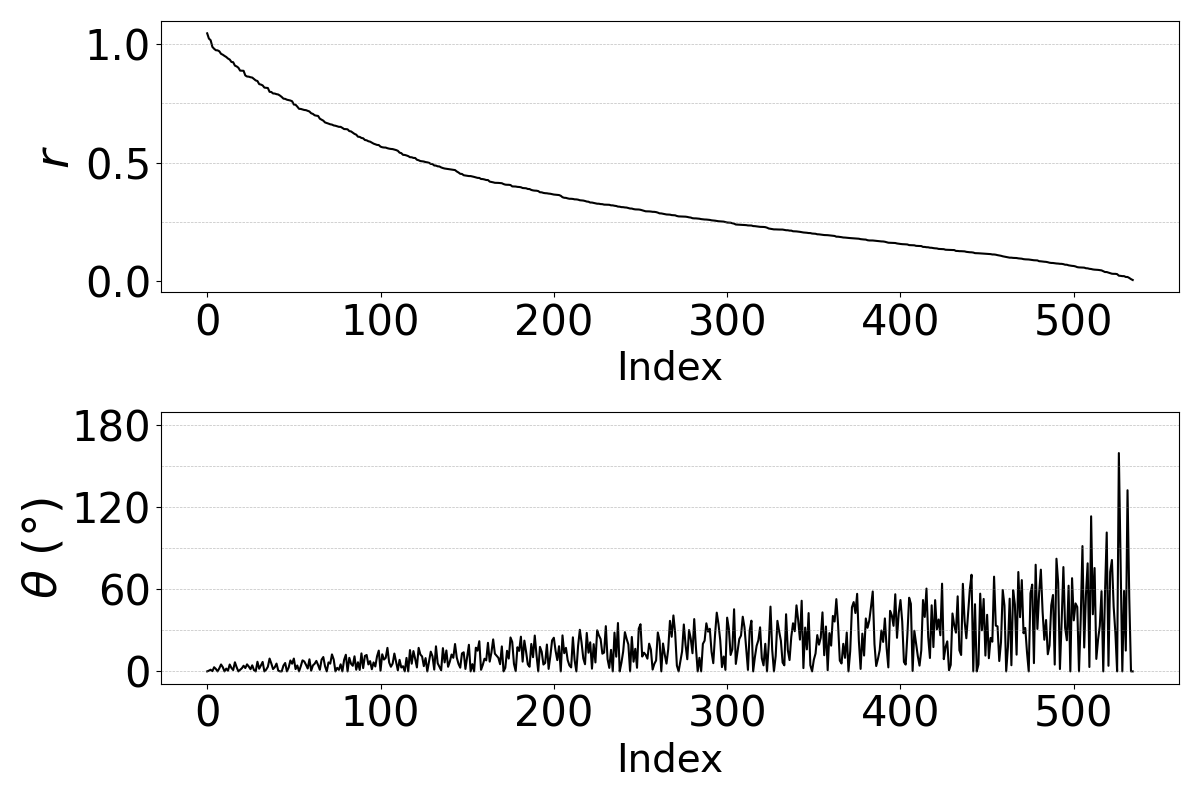}
        (d) clb $\to$ slt
    \end{minipage}
    \caption{Spectra of eigenvalues of the conversion matrix for representative speaker pairs.}
    \label{fig:rotation_spectrum}
\end{figure}

Figure~\ref{fig:angles} showed cosine angles typically in the 25--30$^\circ$ range.
Female$\to$female had the smallest angles, followed by male$\to$male, then cross-gender pairs, with no clear difference between sonorants and obstruents.
Figure~\ref{fig:rotation_spectrum} shows the $(r,\theta)$ spectra of eigenvalues sorted by descending $r$ for the four speaker pairs from subsection~\ref{subsec:sub_eval}.
Planes with high $r$ typically had small rotation angles, while low $r$ showed larger variability but contributed less due to strong shrinkage.
Thus, the transformation is a contractive rotation, concentrating information in few components.
In addition, the clb$\to$slt pair showed smaller rotation angles, consistent with the female$\to$female trend in Figure~\ref{fig:angles}.
This tentatively suggests that $\theta$ may reflect inter-speaker acoustic distance, though this warrants further investigation.

%% file: tex/conclusion.tex
\section{Conclusion}

We presented SSL-GMMVC, a GMM-based voice conversion method that extends a single global transform to locally linear mappings in SSL feature space.
Objective and subjective evaluations showed that SSL-GMMVC improves speaker similarity over LinearVC in particular settings with comparable intelligibility and naturalness.
Even the constrained covariance variant with multiple components surpassed FreeVC in speaker similarity when increasing the number of components.
Further analysis revealed that mixture component selection correlates with phonetic structure, and that the learned transforms are characterized as contractive rotations, with angles tentatively suggesting a relationship to inter-speaker acoustic distance.
However, scaling the number of mixture components while maintaining stable parameter estimation in high-dimensional SSL space remains an open challenge, and the rotation-spectrum analysis does not yet establish correspondences between rotational planes across speaker pairs.
Future work includes addressing these limitations and further exploring the structure of SSL representations through the lens of VC.

%% file: tex/ack.tex
\section{Acknowledgements}
This research was supported by JSPS KAKENHI Grant Number JP25K22829.
The authors would like to thank Kentaro Onda at The University of Tokyo, for his valuable assistance in refining this work.

%% file: tex/ai_use.tex
\section{Use of generative AI tools}
GPT-5.2 was used to aid editing and polishing this manuscript.